\documentclass[10pt]{article}
\usepackage{mathrsfs}
\usepackage{amsmath,amsfonts,amssymb,mathtools}
\usepackage{cite}
\usepackage{dsfont}
\usepackage{graphicx}
\usepackage{hyperref}
\usepackage{verbatim}
\usepackage{slashed}
\usepackage[all]{xy}
\usepackage{xcolor}
\usepackage{subfig}
\usepackage{tikz}
\usetikzlibrary{shapes.geometric,arrows}
\usepackage{relsize}
\usepackage{caption}
\usepackage{float}
\usetikzlibrary{positioning}
\usepackage{geometry}
\usepackage[T1]{fontenc}
\usepackage{lmodern}
\usepackage{stackengine} 
 
\usepackage[utf8]{inputenc}

\hypersetup{
  colorlinks,
  citecolor=violet,
  linkcolor=blue,
  urlcolor=blue}

\csname @addtoreset\endcsname{equation}{section}
\DeclareMathAlphabet\mathbfcal{OMS}{cmsy}{b}{n}

\newcommand{\beq}{\begin{equation}}
\newcommand{\eeq}{\end{equation}}
\newcommand{\bea}{\begin{eqnarray}}
\newcommand{\eea}{\end{eqnarray}}
\newcommand{\ba}{\begin{array}}
\newcommand{\ea}{\end{array}}
\newcommand{\bit}{\begin{itemize}}
\newcommand{\eit}{\end{itemize}}
\newcommand{\nn}{\nonumber}

\newcommand{\complesso}{{\ \hbox{{\rm I}\kern-.6em\hbox{\bf C}}}}
\newcommand{\reale}{{\hbox{{\rm I}\kern-.2em\hbox{\rm R}}}}
\newcommand{\uno}{ \,  \raisebox{+0.14em}{{\hbox{{\rm \scriptsize ]}} \raisebox{-0.2em}{\kern-.8em\hbox{1}}}} \, }  

\newcommand{\p}{\partial}

\newcommand{\g}{\gamma}

\newcommand{\D}{\Delta}

\newcommand{\m}{\mu}

\newcommand{\n}{\nu}
\renewcommand{\r}{\rho}
\newcommand{\s}{\sigma}

\renewcommand{\S}{\Sigma}

\renewcommand{\t}{\theta}

\renewcommand{\c}{\chi}

\newcommand{\om}{\omega}
\newcommand{\Om}{\Omega}

\begin{document}


\begin{titlepage}

\vspace{0.3cm}

\begin{flushright}
$LIFT$--10-3.25
\end{flushright}

\vspace{1.0cm}

\begin{center}
\renewcommand{\thefootnote}{\fnsymbol{footnote}}
\vskip 9mm  
{\Huge \bf Kerr Black Holes in
\vskip 10mm
   an Expanding Bubble}
\vskip 37mm
{\large {Marco Astorino$^{a}$\footnote{marco.astorino@gmail.com}
}}\\

\renewcommand{\thefootnote}{\arabic{footnote}}
\setcounter{footnote}{0}
\vskip 8mm
\vspace{0.2 cm}
{\small \textit{$^{a}$Laboratorio Italiano di Fisica Teorica (LIFT),  \\
Via Archimede 20, I-20129 Milano, Italy}\\
} \vspace{0.2 cm}
%

\end{center}

\vspace{5.1 cm}

\begin{center}
{\bf Abstract}
\end{center}
{An exact and analytical solution, in four-dimensional general relativity, describing a collinear array of an arbitrary number of Kerr black holes inside an expanding bubble of nothing is built, thanks to the inverse scattering technique. Physical properties and thermodynamics of the single Kerr in the bubble are studied. No cosmic strings or struts are present.\\
The binary black hole system displays equilibrium configurations, because the expanding bubble surrounding the black holes balances the mutual gravitational attraction of the two constituents.}

\end{titlepage}

\addtocounter{page}{1}

\newpage


\section{Introduction}
\label{sec:introduction}

It is commonly believed that in four-dimensional pure general relativity, that is the standard Einstein's theory of gravitation governed by the equations $R_{\m\n}=0$, the only analytical black hole solution is represented by the Kerr (or Schwarzschild in the static case) metric\footnote{We are referring to black hole metrics which are completely regular outside the event horizons, that is with no strings, no  curvature nor conical singularities.}. This is certainly true for Minkowskian asymptotics; indeed there are no go theorems implementing this boundary condition. However when the fall-off of the spacetime is not flat, it is possible to build some unexpected generalisations of the Schwarzschild or Kerr black hole. In fact in \cite{bubble} a general procedure to build diagonal metrics describing one or more black holes inside an expanding bubble of nothing has been presented. The binary system composed by two Schwarzschild black holes may admit equilibrium configuration because the gravitational attraction between the two black constituents is balanced by the presence of the expanding bubble of nothing \cite{bubble}. We expect that the addition of the angular momenta to the system could improve the picture and its stability because of the gravitational spin-spin interaction between the two rotating sources; this point will be addressed in section \ref{sec:binary}. Indeed, recently this effect has been proven to be effective in sustaining the equilibrium between two extreme Kerr black holes \cite{swirling-binary}. 
Previous attempts based on the same (general) relativistic repulsive principle proved to be ineffective because they were set in flat asymptotic, see for instance \cite{dietz}.\\    
The scope of this paper is to enrich this picture, in particular considering an arbitrary number of stationary rotating black holes inside the expanding bubble, in section \ref{sec:n-kerr}. A simple system composed by a single rotating black hole in the bubble will be explicitly provided and studied in detail in section \ref{sec:1-kerr}.
Thus, remarkably, general relativity in vacuum admits other single black hole metrics apart from Kerr, Schwarzschild or their immersion in a swirling background \cite{swirling}, \cite{removal}.\\

\section{Brief review of the expanding bubble of nothing}
\label{sec:bubble}

The expanding bubble of nothing, found by Witten \cite{witten} as the final state of the decay of a Kaluza-Klein vacuum, is a regular solution of vacuum Einstein gravity. It represents a surface of spherical topology which contracts till reaching a minimal radius $r_0$ and then expands for increasing time-like coordinate. It is described by the metric
\beq \label{bubble}
      ds^2 = r^2 (-d \c^2 + \cosh^2 \c d\varphi^2) + \left(1-\frac{r_0}{r} \right) d\phi^2 + \frac{dr^2}{1-\frac{r_0}{r}} \ ,
\eeq
which can be obtained from the Schwarzschild metric
\beq\label{sch-bh}
-\left(1-\frac{2m}{r} \right) d\tau^2 + \frac{dr^2}{1-\frac{2m}{r}} + r^2 \left( d\theta^2 + \sin^2 \theta d\varphi^2 \right)   \ ,
\eeq
by two different analytic continuation procedures. The first is by operating on (\ref{sch-bh}) the change of coordinates $\tau = i \phi, \ \theta = i \chi + \pi/2 , \ m = r_0/2$, as given in eq. (\ref{bubble}). The second by a double Wick rotation ($\tau = i\phi , \ \varphi = i t, \ m=r_0/2$) which gives
\beq \label{bubble-dwr}
       ds^2 = \left(1-\frac{r_0}{r} \right) d\phi^2 + \frac{dr^2}{1-\frac{r_0}{r}} + r^2 (d\theta^2-\sin^2\theta dt^2) \ .
\eeq
Clearly the two metrics in eq. (\ref{bubble}) and (\ref{bubble-dwr}) are diffeomorphic: they are related by the coordinates transformation
\beq
      \chi = \arcsin (\sin \theta \sinh t) \ , \hspace{2.3cm} \varphi = \arctan (\cot \theta \ \text{sech} \, t) \ .
\eeq
The three-dimensional $r=r_0$ hyper-surface of the bubble has constant curvature, so in a certain sense it resemble the de Sitter spacetime.
In \cite{bubble}, it has been shown that the bubble can be interpreted, in four dimensions, as the geometry in between two very large black holes. In practice it means that the bubble can be obtained from a double black hole metric, i.e. the Bach-Weyl solution in the limit of large event horizons, while keeping their distance finite \cite{bubble}. For more details about the bubble of nothing see \cite{horowitz-bubble-baths}, \cite{horowitz-maeda} and \cite{bubble}.\\
In Weyl coordinates
\beq
            \r = \sqrt{r^2- r r_0} \, \sin \theta \ , \hspace{2cm} z = z_0 + (r-m) \cos \theta \ .
\eeq
the bubble of nothing, in the form of (\ref{bubble-dwr}), becomes
\beq \label{bubble-weyl}
       ds^2 = - \r^2 \frac{\m_2}{\m_1} dt^2 + \frac{16 \, C_f \, \m_1 \m_2^3}{\m_{12}W_{11}W_{22}} \, (d\r^2+dz^2) + \frac{\m_1}{\m_2} d\phi^2 \ ,
\eeq
where
\beq
             \m_i := w_i - z + \sqrt{\r^2+(z-w_i)^2} \ , \hspace{1.5cm} \m_{ij} := (\m_i-\m_j)^2 \ , \hspace{1.5cm} W_{ij} = \r^2+\m_i\m_j \ .
\eeq
$C_f$ is an arbitrary gauge constant, which is usually fixed to remove a conical defect of the metric, while keeping the usual periodicity of the azimuthal angle $2\pi$. The physical parameters of the solution are encoded into the $w_i$, with $w_i<w_{i+1}$, and to match the metric (\ref{bubble-dwr}) have to be chosen as follows
\beq
      w_1 = z_0 - \frac{r_0}{2} \ ,  \hspace{1.5cm}  w_2 = z_0 + \frac{r_0}{2} \ , \hspace{1.5cm} C_f = - \frac{r_0^2}{4} \ .
\eeq
\\

\section{N Kerr black holes in an Expanding bubble}
\label{sec:n-kerr}

Thanks to the generating technique of Belinski and Sakharov \cite{Belinsky:1979mh}, \cite{belinski-book} we are able to consider the above expanding bubble of nothing as background and embed inside it an arbitrary number, $N$, of collinear rotating black holes of Kerr-NUT type along the $z$-axis. There are several possible equivalent ways to reach this result. The first is to add to the Minkowski background $2N+2$ solitions, two for the bubble and a pair for each rotating black hole. The second consists in using the bubble as background for the inverse scattering and add $2N$ solitions for the black holes only. The third could be implemented starting from a collinear solution of $N+2$ black holes ($2N+4$ solitions on Minkowski) and take the limit for infinitely large event horizon of the external black holes \cite{bubble}. The fourth consist in taking the double Wick rotation of $N+1$ asymptotically flat black holes (so $2N+2$ solitions on a Minkowski background). Lastly, we can start from a collection of $N+1$ accelerating black holes, as described in \cite{multipolar-acc} and blow infinitely the event horizon of the black hole farther with respect to the Rindler horizon.   \\
We will present here the first approach, i.e. by adding $n:=2N+2$ solitions to the flat spacetime, for convenience parametrised as follows
\beq \label{minkowski}
            ds^2 = - \r^2 dt^2 + d\r^2 + dz^2 + d\phi^2 \ .
\eeq
This background can be described in terms of a generic stationary and axisymmetric 
\beq \label{metric-ist}
          ds^2 = g_{ab}(\rho,z) dx^a dx^b + f(\r,z)(d\r^2+dz^2)
\eeq
with $a,b,c \in \{0,1\} $ and $k,l \in \{1,..., n\}$, so $x^a = \{t,\varphi\}$, by the background fields
\beq\label{metric-circ}
        \overset{\circ}{g}_{ab} = \bigg( \begin{tabular}{cc}
       $- \rho^2$ & 0 \\ 
        0 & 1 
        \end{tabular} \bigg)   \hspace{2mm} ,     \hspace{12mm}   \overset{\circ}{f}=1  \ .
\eeq
Then, according to the inverse scattering procedure, the generic metric with $2N+2$ solitons on the background (\ref{minkowski}), (\ref{metric-circ}) can be built in the following way
\begin{subequations}
\label{metric-ph}
\begin{align}
\label{g-ph}
g_{ab}(\rho,z) & =  \frac{1}{\rho^n} \Biggl(\prod_{k=1}^n \mu_k\Biggr) \Biggl[ \ \overset{\circ}{g}_{ab} - \sum_{k,l=1}^n \frac{(\Gamma^{-1})_{kl} L_a^{(k)} L_b^{(l)}}{\mu_k\mu_l}\Biggr] \, , \\
\label{f-ph}
f(\rho,z) & =  \frac{16 \ C_f \overset{\circ}{f}_0}{\rho^{n^2/2}} \ \Biggl( \prod_{k=1}^n \mu_k \Biggr)^{n+1} \ \Biggl[ \prod_{k>l=1}^n (\mu_k-\mu_l)^{-2} \Biggr] \det\Gamma \, ,
\end{align}
\end{subequations}
where $L^{(k)}_a = m^{(k)}_c \overset{\circ}{g}_{ca}$ and
\beq \label{Gamma-mak}
      \Gamma_{kl} = \frac{m_a^{(k)} \  \overset{\circ}{g}_{ab} \ m_b^{(l)}}{\rho^2 + \mu_k \mu_l}\hspace{3mm} , \hspace{12mm}  m^{(k)}_a = \left( \frac{C_0^{(k)}}{\m_k} , C_1^{(k)} \right) \ .
\eeq

$n$ solitons bring into the metric $2n$ physical integration constants, which have to be properly parametrised to obtain the black hole array inside the Witten bubble. Our choice is
\begin{subequations}\label{C01}
\begin{align}
C_0^{(2i)} &= \frac{a_i+b_i}{2} \, , \qquad  C_1^{(2i)} = \frac{-m_i-\s_i}{2} \, , \ \qquad C_0^{(2i+1)} = 1 \, , \hspace{1.8cm} C_1^{(2i+1)} = \frac{-m_i+\s_i}{a_i+b_i} \,  , \ \\
C_0^{(1)} &= 1 \, , \hspace{17mm}  C_1^{(1)} = \frac{-m_B+\s_B}{a_B+b_B}  \, , \qquad C_0^{(2N+2)} = \frac{a_B+b_B}{2} \, , \qquad C_1^{(2N+2)} = \frac{-m_B-\s_B}{2}  \, , \
\end{align}
\end{subequations}
The constants $m_i$, $a_i$, $\ell_i$ are respectively related to the mass, angular momentum and the NUT parameters, with $i\in{1, ..., N}$, however they do not necessary correspond to the $i$-th black hole's conserved charges, unless in some specific cases. We have also defined
$\sigma_i^2 := m_i^2 - a_i^2 + b_i^2$. On the other hand $m_B=z_B, a_B, b_B, \s_B$ are the quantities related to the bubble, corresponding to the first and last solitons of the array $\m_1$ and $\m_{2N+2}$. In case we want a non--rotating bubble we have to set $a_B=0=b_B$, which henceforward will be the default option. 
The ordered poles $w_k$, with $w_k < w_{k+1}$, are taken as follows 
\beq \label{wi}
w_{2} = z_1-\s_1 \, , \ \ w_{3} = z_1+\s_1 \, , \  ... \  
w_{2i} = z_i - \sigma_i \, , \ \
w_{2i+1} = z_i + \sigma_i \, , \  ...  \ 
w_{2N} = z_N - \sigma_N \, , \ \ w_{2N+1} = z_N + \sigma_N \, , \nn
\eeq
while $w_1$ and $w_{2N+2}$ define the position of the bubble. The position of the $i$-th event horizon is located at $\r=0$ and $z_i-\s_i < z < z_i+\s_i$. The interpretation of the metric as an array of collinear $N$ Kerr-NUT black holes on the z--axis, inside an expanding bubble is quite direct. In fact, in the limit where simultaneously $w_1 \to - \infty$ and $w_{2N+2} \to \infty$ (for instance defining $w_{2N+2}=-w_1=z_B$ and taking $z_B \to \infty $) we get the array of $N$ collinear Kerr-NUT black holes on a Minkowski background. Below, simpler explicit examples with one or a couple of black holes inside the bubble will be described in more details. On the other hand, when we make the black holes vanish, through the limit $w_{2i+1} \to w_{2i}$, $\forall \  i \in {1, ..., N}$, we recover the bubble in the form of eq. (\ref{bubble-weyl}).\\
Note that if we take just $w_1 \to - \infty$ (or  alternatively $w_{2N+2} \to \infty$), we obtain an accelerating array of $N$ Kerr-NUT black holes, as in \cite{marcoa-equivalence}.\\

\section{Single Kerr-NUT in the Bubble}
\label{sec:1-kerr}

From the general solution (\ref{metric-ist})-(\ref{Gamma-mak}) we consider $n=4$ solitons to build a single ($N=1$) Kerr-NUT black hole inside and in the center of a expanding,  non-rotating bubble. The relevant physical parameters are chosen as follows
\begin{subequations} \label{C01-single}
\begin{align}
C_0^{(2)} &= \frac{a+b}{2} \, , \hspace{1.2cm}  C_1^{(2)} = \frac{-m-\s}{2} \, , \hspace{1.2cm} C_0^{(3)} = 1 \, , \hspace{1.4cm} C_1^{(3)} = \frac{-m+\s}{a+b} \,  , \ \ \\
C_0^{(1)} &= 1 \, , \hspace{19mm}  C_1^{(1)} = 0  \, , \hspace{16mm}  \qquad C_0^{(4)} = 0 \, , \hspace{1.4cm} C_1^{(4)} = -z_B  \, , \ \ \\
w_1 &= - z_B \ , \hspace{1.6cm}  w_2 = z_{bh} - \s \ , \hspace{1.55cm} w_3 = z_{bh} + \s \ , \hspace{0.65cm} w_4 =  z_B \ . \ \
\end{align}
\end{subequations}
Since we would like to obtain a regular configuration without Misner or cosmic strings, a symmetric configuration has to be chosen. We mean that the black hole is located in the middle of the bubble, along the azimuthal axis, more specifically we restrict the above solution to the case $z_{bh}=0$\ \footnote{The fact that the bubble is parametrised only by the bubble extension $z_B$ instead of two independent parameters ($w_1$ and $w_4$) is not an extra restricting choice.}.  \\
It is not difficult to realise that the spacetime describes one rotating black hole endowed with gravitomagnetic charge enclosed into the bubble of nothing. In fact, by pushing the bubble far away, that is by taking the limit for $z_B \to \infty$, we recover the usual Kerr-NUT black hole, which in Weyl coordinates, takes the form
\beq
          ds^2 = -h (d\bar{t} - \omega d\bar{\varphi})^2 + \frac{1}{h} \left[ e^{2\g} (d\r^2 + dz^2) +\r^2 d\bar{\varphi}^2 \right]\ , 
\eeq
with 
\bea
       h(\r,z)         & = & \frac{a^2(R_+-R_-)^2 + \left[(R_++R_-)^2-4(m^2+b^2)\right]}{a^2(R_+-R_-)^2 - 4ab\s(R_+-R_-)+\left[(R_++R_-+2m)^2+4b^2\right]\s^2} \ ,  \nn \\
       \omega (\r,z)   & = &  \frac{b\s(R_+-R_-)\left[4\s^2-(R_++R_-)^2\right] + a \left[2b^2+m(2m+R_++R_-)\right] \left[4\s^2-(R_+-R_-)^2\right]}{a^2(R_+-R_-)^2} \ , \nn \\
       e^{2\g(\r,z)}   & = &    \frac{\left(b^2+m^2\right) \left[4 \left(b^2+m^2\right)-(R_+ + R_-)^2\right]-4 a^2 \left(b^2+m^2-R_+ R_-\right)}{4 R_+ R_- \sigma ^2}  \ ,  \nn   \\
       R_\pm (\r,z)    & = & \sqrt{\r^2 + (z \pm \s)^2}\ ,
\eea
where before taking the limit, we have rescaled the time and the azimuthal angle such as 
\beq \label{rescaling}
         t =  \frac{\bar{t}}{2 z_B} \ ,  \hspace{2cm} \varphi = \bar{\varphi} \, (2 z_B) \ .
\eeq
While usual Kerr-NUT metric in spherical coordinates ($ t, r, x:= \cos\theta, \varphi$)
\beq
      ds^2 = - \frac{\D_r}{\S^2} \left[d\bar{t} + \left(2b\cos \theta +a \sin^2 \theta \right) d\bar{\varphi}\right]^2 +\S^2 \left( \frac{dr^2}{\D_r} + d\theta^2 \right) +\frac{\sin^2 \theta}{\S^2} \left[ad\bar{t} + (r^2+a^2+b^2) d\bar{\varphi}\right]^2 \ , 
\eeq
with
\bea
      \D_r(r) = (r-m)^2-\s^2 \ ,  \hspace{1.4cm}  \S^2(r,\theta) = r^2 + (a\cos\theta-b)^2 \ \ ,
\eea
is retrieved thanks to the coordinates transformation
\beq
              \r(r,\theta) = \sqrt{\D_r} \, \sin \theta \ , \hspace{2cm} z(r,\theta) = (r-m) \cos \theta \ .
\eeq
On the other hand, to obtain only the bubble of nothing the limit $w_2 \to w_3$ has to be taken. In terms of the parametrization (\ref{C01-single}) it means that $\s \to 0$. However, imposing only the vanishing of $\s$ is not sufficient, because it determines the extremal Kerr-NUT configuration. Thus, the correct limiting procedure to get only the bubble is to vanish simultaneously all the physical parameters of the black hole, that is $m=b=a=0$, which gives $\s=0$ also outside extremality. \\

\paragraph{Regularisation} Our main goal is to build a solution without mathematical or physical criticalities, at least outside the event horizon. In particular we remember that the presence of the Misner string also generates curvature singularities on the z--axis, more harming than a conical one, because these can not be absorbed in gauge transformations. Thus its removal is a necessary condition for the manifold to be well defined. Therefore, first of all, we start to identify which are the parameters related to the presence of the NUT parameter, to remove them from the spacetime. The Misner string is a delta-like angular momentum singularity on the symmetry axis, which can be detected inspecting the value of the $\omega = - g_{t\varphi}/g_{tt} $ function, in the domain of outer communication of the black hole. In this case we have to evaluate in the two segments, on the z-axis, between the black hole horizon and the bubble border the following quantity
\beq
           \Delta \omega = \omega_+ -\omega_- \ , 
\eeq   
where 
\bea
      \omega_+ &=&  \lim_{\r \to 0} \left(- \frac{g_{t\varphi}}{g_{tt}} \right) \hspace{1cm} \textrm{for} \hspace{1cm} -z_B \leq z \leq -\s \ ,\\
      \omega_{-} &=&  \lim_{\r \to 0} \left(- \frac{g_{t\varphi}}{g_{tt}} \right)  \hspace{1cm} \textrm{for} \hspace{1.6cm} \s \leq z \leq z_B \ .
 \eea
It is not difficult to realise that rotation of the azimuthal axis on the two disconnected regions is the same constant, hence $\D \omega =0$, for $b=0$. Then this value, characterising the angular velocity of the axis, can be set to zero by a linear coordinate transformation between the time and the azimuthal angle coordinates, which introduce a constant shift, $\omega_0$, in the $\omega(\r,z)$ function. This gauge constant that usually defines the angular velocity of an asymptotic observer can be set to guarantee that there is no source of angular momentum on the z-axis. In particular, in the sub-case of the single rotating black hole in the bubble, we require this gauge constant to be 
\beq \label{w0}
         \omega_0 = \frac{a}{2(z_B^2-\s^2)} \ \ .
\eeq
So the NUT parameter is exactly related to the $b$ constant, as for the asymptotically flat Kerr black hole, that's because in this section, in contrast with the general solution of section \ref{sec:n-kerr}, we have imposed no rotational parameter, that is no NUT nor angular momentum on the bubble. Otherwise the general NUT parameter of the black hole would be related also to these other physical constants.\\
Once the Misner strings are removed, we can analyse possible axial conical singularities, which, from a physical point of view, indicate the presence of cosmic strings or struts. Their presence is identified by considering the ratio between a small circle around the two black hole poles\footnote{The gauge choice in eq. (\ref{w0}) implies a shift of the $g_{\varphi\varphi}$ as in (\ref{g-ph}): $g_{\varphi\varphi} \leadsto \tilde{g}_{\varphi\varphi} = g_{\varphi\varphi}-2\omega_0g_{t\varphi}+g_{tt}\om_0^2$ and $g_{t\varphi} \leadsto \tilde{g}_{t\varphi} = g_{\t\varphi}-\omega_0g_{tt}$, while $ f = g_{\r\r} = g_{zz} $ and $g_{tt}$ remain unvaried.} $L=2\pi\sqrt{\tilde{g}_{\varphi\varphi}}$ and  their radius $R=\r\sqrt{g_{zz}}$, that is
\beq \label{ratio}
             \lim_{\rho \to 0} \frac{L}{R} \ = \ \lim_{\r \to 0} \frac{2\pi}{\r} \sqrt{\frac{\tilde{g}_{\varphi\varphi}}{g_{zz}}} \ .
\eeq
When this quantity is $2\pi$, outside the event horizon, the spacetime is void of conical singularities. On the contrary, the metric is plagued by angular deficit or excess whenever the above ratio is smaller or bigger than $2\pi$. Since we intelligently have chosen a symmetric configuration, where $z_{bh}=0$, the values of (\ref{ratio}) on the two antipodal sectors of the event horizon coincide. So it is easy to exploit the residual gauge freedom encoded in the constant $C_f$ to regularise the metric. Specifically we can choose
\beq \label{Cf}
         C_f = 16 \frac{(z_B^2-\s^2)^2}{z_B^2} \ 
\eeq
to be sure that the spacetime is not plagued by cosmic strings or struts. \\
After selecting  $\omega_0$ and $C_f$ as in eqs. (\ref{w0}) and (\ref{Cf}) respectively, the only remaining singularity on this metric is the curvature singularity enclosed inside the black hole, as in the standard black hole models. Therefore this spacetime legitimately describes a Kerr black hole inside the expanding bubble of nothing. This solution generalises the one presented in \cite{bubble} because of the presence of the angular momentum of the black hole. Further generalisations can be considered following section \ref{sec:n-kerr} by keeping non--zero the rotating parameters, also for the bubble of nothing. Further electromagnetic generalisations of black hole in the expanding bubble can be found in \cite{bubbleverse}.   \\

\paragraph{Charges and Thermodynamics} We are interested in computing some physical and thermodynamic properties of the black hole inside the bubble. Here we are mainly interested into specific quantities which characterise the black hole, possibly indirectly influenced by the presence of the bubble, but we are not directly taking into consideration bubble quantities, which may diverge since they extend towards infinity. Thus integrands will be evaluated excluding the portion of the manifold represented by the bubble. \\
The mass of the black hole can be computed by the Komar integral 
\beq
           M = \frac{1}{8 \pi}\int_{0}^{2\pi} d\varphi \int_{-z_B}^{z_B} \r \left( \tilde{g}^{tt} \p_\r \tilde{g}_{tt} + \tilde{g}^{t\varphi} \p_\r \tilde{g}_{t\varphi}  \right) dz \ \bigg|_{\r=0} \  = \ m  \ . 
\eeq 
Similarly the Komar angular momentum 
\beq
           J =  -\frac{1}{16 \pi} \int_{0}^{2\pi} d\varphi \int_{-z_B}^{z_B}   \r \left( \tilde{g}^{tt} \p_\r \tilde{g}_{t\varphi} + \tilde{g}^{t\varphi} \p_\r \tilde{g}_{\varphi\varphi}  \right) dz \ \bigg|_{\r=0} \   = \frac{am}{4(a^2-m^2+z_B^2)}   \ . 
\eeq 
Note that the integrals vanish outside the event horizon (i.e. for $-z_B<z<-\s$ and $\s<z<z_B$). \\
The area of the event horizon is given by
\beq
        A = \int_{0}^{2\pi} d\varphi \int_{-\s}^{\s} \sqrt{f \ g_{\varphi \varphi}} \ dz \ = \ \frac{2 \pi m r_+}{z_B^2+a^2-m^2} \ . 
\eeq
The black hole temperature computed from the surface gravity or by the regularisation of euclidean continuation of the metric is
\beq
        T = \frac{- \tilde{g}_{tt}}{2 \pi \tilde{g}_{t\varphi} \sqrt{f \ \tilde{g}_{tt}} } \ \bigg|_{\r=0}
\ = \ \frac{\s (z_B^2-\s^2)}{\pi m r_+} \ . \eeq
The angular velocity of the horizon is 
\beq
         \Om = - \frac{\tilde{g}_{t\varphi}}{\tilde{g}_{\varphi\varphi}} = \frac{2a(z_B^2-\s^2)}{m r_+} \ .
\eeq
The physical quantities of the usual Kerr black hole are retrieved when the bubble is pushed at spatial infinity for $z_B \to \infty$. In that limit the time and azimuthal angle rescaling (\ref{rescaling}) has to be taken into consideration. \\
Employing the Bekenstein-Hawking value for the black hole entropy, as a quarter of the event horizon area: $S=A/4$, the Smarr law can be easily verified
\beq
         M = T  S + 2 \Om  J \ .
\eeq
On the other hand, the first law of black hole thermodynamics 
is not working properly for this non-trivial background. Possibly this inconvenient can be solved by considering the contribution of the expanding bubble or by the introducing a proper normalisation for the time-like Killing vector, as explained in \cite{kerr-Melvin-mass} and successfully applied in several non-trivial asymptotic conditions, such as magnetised or accelerating black holes \cite{kerr-Melvin-mass}, \cite{thermo-acc}.  \\

\paragraph{Schwarzschild in the Bubble}

From the above Kerr-NUT black hole inside the bubble it is straightforward to vanish the rotating parameters, i.e. the angular momentum for unit mass $a$ and the NUT parameter $b$, to get the Schwarzschild black hole in the bubble, found in \cite{bubble}
\beq \label{sch+bubble}
              ds^2 = - \frac{\r^2 \m_2 \m_4}{\m_1\m_3} dt^2 + \frac{16 \, C_f \, w_2^2\, w_4^2 \left[C_0^{(1)} C_0^{(3)} C_1^{(2)}C_1^{(4)} \right]^2 \m_1^3 \m_2^5 \m_3^3 \m_4^5}{w_1^2 w_3^2 W_{11} W_{22} W_{33} W_{44}  W_{13}^2 W_{24}^2 \m_{12} \m_{14} \m_{23} \m_{34}} \, (d\r^2+dz^2) + \frac{\m_1\m_3}{\m_2 \m_4}  d\varphi^2 \ .
\eeq
This metric is naturally free of Misner string, while the regularity conditions to avoid cosmic strings or axial struts are given by
\beq \label{Cf-w4}
              C_f = \frac{4 w_1^2 w_3^2 (w_1-w_3)^2 (w_2-w_3)^2 (w_3-w_4)^2}{w_2^2 w_4^2\left[C_0^{(1)} C_0^{(3)} C_1^{(2)}C_1^{(4)} \right]^2} \ \ , \hspace{1.3cm} w_4=-w_1+w_2+w_3 \ .
\eeq
which is compatible with the constraints for the rotating case above. In fact, using relations (\ref{C01-single}) and $z_{bh}=0$, the second constraint of (\ref{Cf-w4}) is satisfied while the first becomes (\ref{Cf}).\\

\begin{figure}[h!]
	\centering
	\includegraphics[scale=0.57]{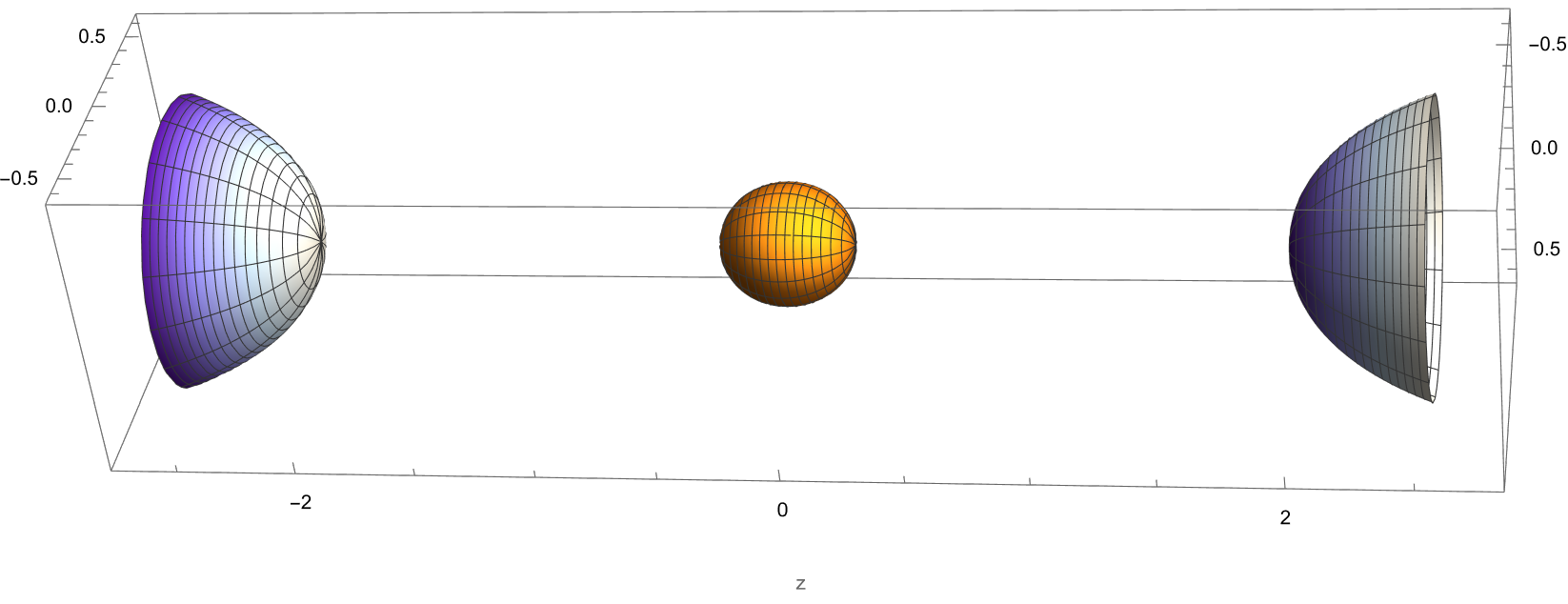}
	\caption{\small Isometric embedding of the Schwarzschild black hole horizon (orange) and (part of ) the  bubble's horizon of nothing (violet) immersed in the three-dimensional Euclidean space; for $ w_1=-4 , w_2=-1, w_3=1, w_4=4$. The geometry of the black hole is deformed by the presence of the surrounding bubble.}
	\label{fig:embedding-schwarzschild}
\end{figure}

An isometric embedding of the spacetime's horizons described by eq (\ref{sch+bubble}) can be found in picture \ref{fig:embedding-schwarzschild} 

The gravitational field of the bubble affects the black hole horizon, which is not perfectly spherical as the Schwarzschild one (when $z_B \to \infty$). From the computation of the  equatorial
\beq
       C_{eq} =  \int_0^{2\pi} \sqrt{g_{\varphi\varphi}} \,d\varphi =  \frac{2\pi m}{z_B}
\eeq

and polar circumference\footnote{Expressed in terms of the complete elliptic integral function.}
\beq
 C_{pol} = 2 \int_{-m}^{m} \sqrt{g_{zz}} \, dz =  \frac{4m \, \textrm{EllipticE}\left(\frac{m^2}{z_B^2}\right)}{z_B^2-m^2} \ ,
\eeq

of the black hole event horizon, we note that $C_{pol}>C_{eq}$; therefore horizon is oblate along the z-axis, as can be appreciated also from Figure \ref{fig:embedding-schwarzschild}.\\
The metric (\ref{sch+bubble}), written in terms of the Belinski-Sakharov solitons, is a little obscure, but it can be greatly simplified passing to spherical coordinates ($\tau,r,\theta,\phi$). Apart trivial constant rescaling of the Killing coordinates, the change of coordinates is defined by 
\bea
       \rho(r,\theta) & = & \frac{\sqrt{(r^2-2mr+B^2m^2r^2)(1-B^2r^2)\sin^2\theta(1-B^2m^2\cos^2 \theta)}}{1+B^2[-r^2+(r^2 - 2mr+B^2m^2r^2)\cos^2 \theta]} \ , \\
          z(r,\theta) & = &   \frac{\cos\theta (r-m)(1-B^2mr)}{1+B^2[-r^2+(r^2 - 2mr+B^2m^2r^2)\cos^2 \theta]}  \ .
\eea
The physical parameters, satisfying the regularity condition in (\ref{Cf-w4}), have been chosen as follows
\beq
           w_1 = - \frac{1}{B} \ , \qquad w_2 = - m  \ , \qquad w_3 = m \ , \qquad w_4 = \frac{1}{B} = z_B \ .
\eeq 
Then the final black hole in the expanding bubble (\ref{sch+bubble}) in spherical coordinates takes the form\footnote{The constant $\D_\phi$ as been introduced by a $\varphi$ rescaling, to remove string or struts, instead of the constant $C_f$.}
\beq  \label{bh-dwr}
         ds^2 = \frac{(1-B^2mr\cos^2\theta +\Omega)^2}{4\, \Om^4} \left[- f(r) d\tau^2 + \frac{dr^2}{f(r)} + \frac{r^2 d\theta^2}{1-B^2m^2\cos^2\theta} \right] + \frac{4 r^2 \sin^2 \theta \, (1-B^2m^2\cos^2\theta) \, \D_\phi^2 \, d\phi^2}{(1-B^2mr\cos^2\theta +\Om)^2}  ,
\eeq
with 
\bea
        f(r) &=& \left(1-\frac{2m}{r} + B^2m^2\right)(1-B^2r^2) \ , \nn \\
        \Om(r,\theta) &=& \sqrt{1-B^2r[r+(2m-B^2m^2r-r)\cos^2\theta]} \ . \nn
\eea
In these coordinates the bubble and the event horizon are located respectively at 
\beq
                r_B = \frac{1}{B} \ , \hspace{2cm}                   r_+ = \frac{2m}{1+B^2m^2} \ .
\eeq
The metric is void of conical singularity for $\D_\phi=1/(1-B^2m^2)$, and the position of the poles $w_3<w_4$ is respected  when $|B|<1/m$.\\
The limit for $B=0$ is well defined and the metric, when the bubble is pushed to infinity disappears, so (\ref{bh-dwr}) reduces to the Schwarzschild black hole in spherical coordinates (\ref{sch-bh}). On the other hand, when the mass parameter $m$ vanishes in (\ref{bh-dwr}), thanks to the following change of coordinates
\beq
        r \to \frac{r_0\sqrt{4\bar{r}^2-4r_0 \bar{r}+r_0^2 \cos^2 \bar{\theta}}}{2\bar{r} - r_0} \ , \qquad \theta \to \arccos \left( \frac{r_0 \cos \bar{\theta}}{\sqrt{4\bar{r}^2-4r_0 \bar{r}+r_0^2 \cos^2 \bar{\theta}}} \right) \ , \qquad t \to r_0  \bar{t} \ , \qquad \phi \to \frac{\bar{\phi}}{2r_0} \ ,  \nn
\eeq
and redefinition of the parameters
\beq
             B \to \frac{1}{r_0} \ \ , \qquad \qquad \D_\varphi \to 1 \ ,
\eeq
we recover the double Wick rotation of the Schwarzschild black hole, as in (\ref{bubble-dwr}), i.e. the expanding bubble of nothing in spherical coordinates $(\bar{t}, \bar{r}, \bar{\theta}, \bar{\phi})$.\\
Note that the analytic continuation of $B \to i B$ remove the bubble of nothing and give rise to a static type I black hole endowed with a gravitational {\it hair} encoded into the integration constant $B$.\\

\section{Rotating Binary black hole system balanced by the Bubble}
\label{sec:binary}

From the general $N$ black hole inside the bubble of nothing in eqs. (\ref{metric-ist})-(\ref{Gamma-mak}) we can restrict to another interesting specific subclass, which describes a binary black hole system, for $N=2$, constituted by $n=6$ solitons. Also in the rotating case we can have equilibrium configuration for a black hole pair, where the gravitational attraction between the collapsed stars is balanced by the presence of the expanding bubble. It is reasonable to think an equilibrium configuration is possible because in the non rotating case the system can be balanced, as shown in \cite{bubble}. However, in the presence of angular momentum, the repulsive effect due to the spin-spin interaction between the two black holes should improve the equilibrium\footnote{Very recently, in \cite{swirling-binary}, it as been shown that the spin-spin interaction could be sufficient to balance the gravitational black hole attraction of a binary system.} and possibly the stability. \\
In particular, in this scope, we can consider a very symmetric case where the bubble is centred in the origin of coordinates and also the black holes are symmetrical with respect to the spatial plane defined by $z=0$. Moreover the two bodies, have the same mass and angular momentum, that is
\beq
w_1=-z_B , \ \ \ w_2=-z_{bh} - \s , \ \ \ w_3 = -z_{bh} + \s , \ \ \ w_4= z_{bh} - \s , \ \ \ w_5 = z_{bh} + \s  , \ \ \ w_6=z_B \ ,
\eeq
with $\s<z_{bh}$, $z_b > z_{bh}+\s $ and
\beq
        \s_1 = \s_2 = \s , \quad m_1 = m_2 = m , \quad a_1 = a_2 = a , \quad b_1 = b_2 = b \ . \quad
\eeq
Thanks to this minimal parametric configuration, we are able to carry on the computations about possible equilibrium configurations. Specifically, we analyse the presence for angular defects on the $z$-axis similarly to the single black hole case in section \ref{sec:1-kerr}.
To remove the conical singularity, it is necessary firstly to remove the Misner string from the three sectors of the metric between the black sources, i.e. $(-z_B,-z_{bh}-\s),(-z_{bh}+\s,z_{bh}-\s)$ and $(z_{bh}+\s,z_B)$. In this scope, it is sufficient to fix $b=0$, $a$ and the $\om_0$ parameter, which determines the rotational angular velocity, to be zero on the axis of symmetry outside the black holes and bubble horizons. \\
Then the conical singularities are avoided, by imposing (\ref{ratio}) $= 2\pi$ on the same above three sectors, just by setting the remaining unconstrained physical parameters, $C_f$ and $m$. Note that, because of the symmetric configuration, fixing two parameters is sufficient to regularise all the three sectors. The computations have been carried out in an analytical and exact way, but with fixing for simplicity $z_B$ and $z_{bh}$ to arbitrary numerical values. \\




\section{Conclusions}

The Kerr solution can be generalised to an expanding bubble of nothing as background. Thanks to the solution generating technique, the metric has been built analytically. It is not singular outside the event horizon, therefore it can be considered a legitimate stationary extension of the rotating black hole. \\
Actually we can build a collection of collinear Kerr-NUT black holes enclosed into the Witten bubble. The presence of the bubble allows us to keep at equilibrium both a single or a rotating binary black hole system. The analogy of the bubble with the de Sitter spacetime, such as the constant curvature of the bubble or the fact that it acts as a couple of accelerating horizons, hints that an analytical and exact metric describing a binary black hole into de Sitter can be built. This is in line with the recent numerical results obtained in \cite{Dias:2024dxg}, where it is said that the presence of the rotation may generate a repulsive force, which helps the stability of the system\footnote{Similarly in the presence of massive complex scalar field numerical results \cite{Herdeiro:2023roz} indicate that the equilibrium between two spinning black hole is possible also for flat asymptotic.}. \\
For future perspective we mention that it is not difficult to generate charged generalizations of these metrics, thanks to the inverse scattering technique \cite{belinski-book}\footnote{The use of the Harrison transformation to charge these configuration would not generate charged black holes in the bubble, because the background would not be electromagnetically neutral. On the other hand the inverse scattering allows one to remain only with a Witten bubble as a background.}. \\
We conclude by noticing  that while these solutions are clearly different from the asymptotically flat counterpart, the spirit of the no-hair theorems still holds, since they still represent Kerr black holes, just in a different background. The deformation introduced by the extra gravitational field is continuously connected to the usual Kerr black hole and can be made, for phenomenological purpose as small as needed, just  by enlarging the dimension of the bubble, via its characteristic parameter $z_B$.\\

\vspace{0.1 cm}

\paragraph{Acknowledgements}
{\small I would like to thank Adriano Viganò for valuable discussions on this subject. A Mathematica notebook containing the main solution presented in this article can be found in the arXiv source folder.}\\
\vspace{1.8cm}

\begin{figure}[h!]
	\centering
	\includegraphics[scale=1]{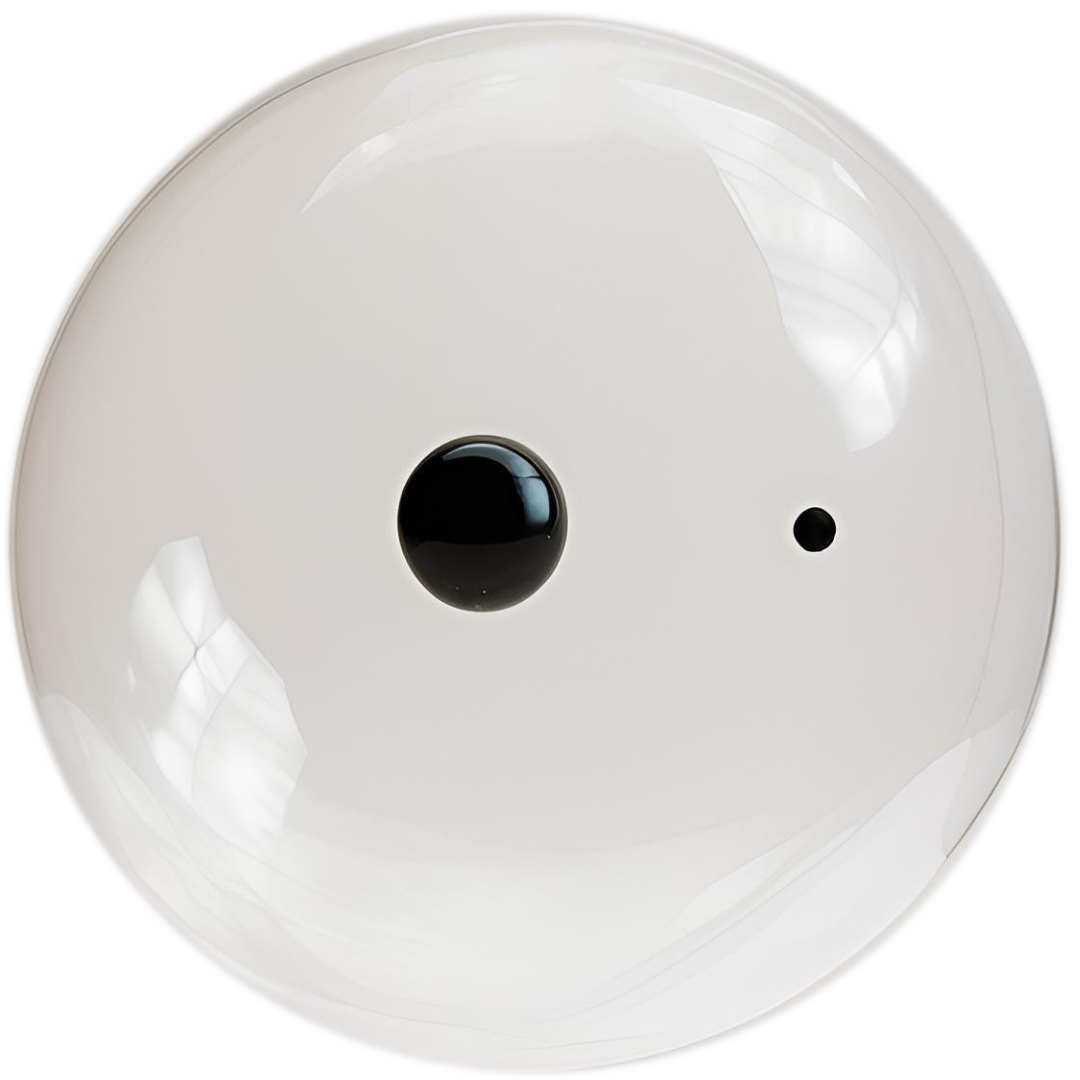}
	\caption{\small A pictorial illustration representing a couple of collinear black hole enclosed in a bubble. See \ref{fig:embedding-schwarzschild}, for a more precise image based on the actual metric solution, specifically the Schwarzschild black hole horizon in the bubble of nothing embedded in the three-dimensional Euclidean space.}
	\label{fig:horizons}
\end{figure}



\end{document}